\documentclass[conference, letterpaper]{IEEEtran}
\ifCLASSINFOpdf
\else
\fi

\usepackage{subcaption}
\ifCLASSINFOpdf 
   \usepackage[pdftex]{graphicx}
\else
\fi

\usepackage{amsfonts}
\usepackage{amssymb}
\usepackage[mathscr]{euscript}
\usepackage{stmaryrd} 
\usepackage{hyperref} 
\usepackage{graphics} 
\usepackage{bbm} 
\usepackage{dblfloatfix}

\usepackage{tikz}
\usepackage{framed}

\usepackage[cmex10]{amsmath} 
\usepackage{color}
\usepackage{fancyhdr}

\renewcommand{\thispagestyle}[2]{} 
\fancypagestyle{plain}{
        \fancyhead{}
        \fancyhead[C]{first page center header}
        \fancyfoot{}
        \fancyfoot[C]{first page center footer}
}
\begin{document}

%
\title{Modeling Smart Contracts Activities:\\
A Tensor Based Approach}

\author{\IEEEauthorblockN{Jeremy Charlier}
\IEEEauthorblockA{Security and Trust Center (SnT)\\
University of Luxembourg\\
Luxembourg, Luxembourg\\
Email: jeremy.charlier@uni.lu}
\and
\IEEEauthorblockN{Radu State}
\IEEEauthorblockA{Security and Trust Center (SnT)\\
University of Luxembourg\\
Luxembourg, Luxembourg\\
Email: radu.state@uni.lu}
\and
\IEEEauthorblockN{Jean Hilger}
\IEEEauthorblockA{Information Technology\\
BCEE\\
Luxembourg, Luxembourg\\
Email: j.hilger@bcee.lu}}


%


\maketitle

\begin{abstract}
Smart contracts are autonomous software executing predefined conditions. 
Two of the biggest advantages of the smart contracts are secured protocols and transaction costs reduction. 
On the Ethereum platform, an open-source blockchain-based platform, smart contracts implement a distributed virtual machine on the distributed ledger. 
To avoid denial of service attacks and monetize the services, payment transactions are executed whenever code is being executed between contracts. 
It is thus natural to investigate if predictive analysis is capable to forecast these interactions. 
We have addressed this issue and propose an innovative application of the tensor decomposition CANDECOMP/PARAFAC to the temporal link prediction of smart contracts. 
We introduce a  new approach leveraging  stochastic processes for series predictions based on the tensor decomposition that can be used for smart contracts predictive analytics.
\end{abstract}


\IEEEpeerreviewmaketitle

\section{INTRODUCTION}
With more and more financial and IoT specific applications being implemented on top of distributed ledgers and associated monetization realized with several crypto-currencies, the modeling and predictive analytics of smart contracts is essential for multiple cases. 
Anti Money Laundering (AML) compliance checking is becoming mandatory and novel investment products do need a framework for modeling and analyzing smart contracts. 
The Ethereum platform has already more than one million accounts with little support existing in the literature on modeling and predicting the interactions among them. 
We have thus addressed the modeling and predictive analytics of the interactions among smart contracts from a multi-disciplinary viewpoint. 
We propose a multi-dimensional decomposition technique leveraging multi-dimensional tensors for extracting relevant latent factors and rely on specific time series models used in the financial industry associated to advanced calibration and Monte Carlo simulations. 
In order to describe our approach, we will first give a fast introduction to smart contracts and tensor models in the section 1 of the paper. 
Section 2 provides  the fundamentals of tensor decomposition and, in section 3, we describe the stochastic model used for the smart contracts activities prediction. 
We report experimental  results on a large dataset in section 4 and  address a final conclusion and pointers to future works in the last section.

The main contribution of this paper consists in a  tensor modeling approach for smart contracts. 
A second contribution is the prediction of smart contracts activities with a geometric Brownian motion combined with a Ornstein-Uhlenbeck process. %

\subsection{Smart Contracts Background}
The computer scientist, Nick Szabo, introduced in 1994 the expression smart contracts as "\textit{a computerized transaction protocol that executes the terms of a contract [...] to satisfy common contractual conditions, minimize exceptions [and] the need for trusted intermediaries. 
Related economic goals include lowering [...] transaction costs}".
Smart contracts have found a direct application in the Ethereum platform that allows every programmer to create their smart contracts to send crypto token. 
Ethereum claimed transparent transaction and execution through a democratic organization which ensures more stability than a central gatekeeper. 
More particularly, in \cite{c1}, Morabito describes how entities can leverage on smart contracts for automate transactions and cost reduction. 
Smart contracts are presented as an efficient way of gaining competitive advantage. 
Swan in \cite{c2} proposes a solution to execute smart contracts under optimal time condition linked to time specifiability. 
This condition is directly implemented in the code of smart contracts for automatic execution. 
Other evolution and programming features arrived such as logic-based programmation for smart contracts. 
In \cite{c3}, the authors proposed logic-based algorithms for further efficiency of the logic approach applied to economic rule.

As illustrated, most of the research is currently focusing on smart contracts optimization or on the legal constraints arising with their use as done in \cite{c4} but not on their activities modeling. 
In our approach, we propose to focus on the analysis of the interactions between smart contracts. 
Moreover, using tensor decomposition and stochastic processes, the objective is to retrieve significant smart contracts activities that will be simulated over time. 

\subsection{Tensor Decomposition Applied To Smart Contracts}
Tensors have appeared as a reliable technique for modeling interactions in multi-dimensional spaces after the introduction of CANDECOMP/PARAFAC (CP) decomposition by Harshman, Caroll and Chang in \cite{c5} and \cite{c6}. 
The ease of the results treatment is one of the main advantages of the CP decomposition. 
It has been widely used in different studies and has been followed by other techniques presented in the extensive survey done by Kolda and Bader in \cite{c7}. 
The tensor theory can be applied from crime forecasting in New York city in \cite{c8} to international trade exchanges in \cite{c9}. 
The authors in \cite{c10} have showed CP decomposition offers good accuracy for time prediction when applied to noisy data. 
This evolution is joined by the development of tensor libraries in Python \cite{c11} as described by Kossaifi, Panagakis and Pantic.
Futhermore, latest research focus on tensor scalability for their use in big data environment as shown by Kijung Shin, Lee Sael and U Kang in \cite{c12}.

As illustrated by the published papers, tensors seem sufficiently versatile to be applied to smart contracts interaction analysis and forecasting activities. 
In addition, all the papers underline good accuracy of experiments results. 
However, papers have not yet proposed a method to model smart contracts interactions using a tensor approach.
The CP tensor decomposition is applied on smart contracts executed on Ethereum platform which are available to all public users for transparency reasons. 

\section{TENSOR DECOMPOSITION}
In this section, we briefly describe mathematical operations involved in CP tensor decomposition before presenting the non-negative CP algorithm used for the analysis.

\subsection{Tensor Description}
\textbf{Notation} Terminology in this paper is very close to the one proposed by Kolda and Bader in \cite{c7} and commonly used by previous publications. Scalars are identified by lower case letters, \textit{a}. Vectors and matrices are denoted by boldface lowercase letters and boldface capital letters, respectively \textbf{a} and \textbf{A}. High order tensors use Euler script notation as $\mathscr{X}$.

\textbf{Tensor Definition} Define $\mathscr{X}\in \mathbb{R}^{I_1 \times I_2 \times I_3 \times ... \times I_n}$ as a \textit{n}-th multidimensional array. $\mathscr{X}$ is called a tensor of order \textit{n}.

\textbf{Tensor Operations} The norm of a tensor $\mathscr{X}$ is defined as the square root of the sum of all tensor entries squared.
\begin{equation} \label{eq::norm}
||\mathscr{X}||=\sqrt{\sum_{j=1}^{I_1}\sum_{j=2}^{I_2}...\sum_{j=n}^{I_n}x_{j_1, j_2, ..., j_n}^2}
\end{equation}
The rank-\textit{R} of a tensor $\mathscr{X}\in\mathbb{R}^{I_1\times I_2\times ...\times I_N}$ is the number of linear components that could fit $\mathscr{X}$ exactly such that
\begin{equation} \label{eq::rank}
\mathscr{X}=\sum_{r=1}^R \textbf{a}_r^{(1)} \circ \textbf{a}_r^{(2)} \circ ... \circ \textbf{a}_r^{(N)}
\end{equation}
with the symbol $\circ$ representing the vector outer product.\\
Matricization, also commonly known as unfolding or flattening, consists in the transformation of a \textit{N}-way array into a matrix. The mode-\textit{n} matricization of the tensor $\mathscr{X}\in \mathbb{R}^{I_1\times I_2\times ...\times I_N}$, denoted \textbf{X}$_{(n)}$, is defined as 
\begin{equation} \label{eq::matricization}
j=1+\sum_{\substack{k=1 \\ k\neq n}}^N(i_k-1)J_k \quad\textup{with}\quad J_k=\sum_{\substack{m=1 \\ m\neq n}}^{k-1}I_m
\end{equation}
where the $(i_1,...,i_d)$ tensor element is mapped to $(i_n,j)$	matrix element.\\
The \textit{n}-mode product of a tensor $\mathscr{X}\in\mathbb{R}^{I_1\times I_2\times ...\times I_N}$ with a matrix $\textbf{M}\in \mathbb{R}^{J\times I_n}$, denoted $\mathscr{X}\times_n \textbf{M}$ results in a tensor of size $I_1\times...\times I_{n-1}\times J\times I_{n+1}\times...\times I_N$. The \textit{n}-mode product is defined by the following equation
\begin{equation} \label{eq::nmode_prod}
(\mathscr{X}\times_n \textbf{M})_{i_1...i_{n-1}ji_{n+1}i_N}=\sum_{i_n=1}^{I_n}x_{i_1i_2...i_n}m_{ji_n}
\end{equation}
The Kronecker product between two matrices \textbf{A}$\in\mathbb{R}^{I\times J}$ and \textbf{B}$\in\mathbb{R}^{K\times L}$, denoted by \textbf{A}$\otimes$\textbf{B}, results in a matrix \textbf{C}$\in\mathbb{R}^{IK\times KL}$.
\begin{equation} \label{eq::kron}
\textbf{C}=\textbf{A}\otimes\textbf{B}=
\begin{bmatrix}
 a_{11}\textbf{B}& a_{12}\textbf{B} & \cdots & a_{1J}\textbf{B}\\ 
 a_{21}\textbf{B}& a_{22}\textbf{B} & \cdots & a_{2J}\textbf{B}\\
 \vdots & \vdots & \ddots & \vdots \\ 
 a_{I1}\textbf{B}& a_{I2}\textbf{B} & \cdots & a_{IJ}\textbf{B}
\end{bmatrix}
\end{equation}
The Khatri-Rao product between two matrices \textbf{A}$\in\mathbb{R}^{I\times K}$ and \textbf{B}$\in\mathbb{R}^{J\times K}$, denoted by \textbf{A}$\odot$\textbf{B}, results in a matrix \textbf{C} of size $\mathbb{R}^{IJ\times K}$. It is the column-wise Kronecker product.
\begin{equation} \label{eq::kr}
\textbf{C}=\textbf{A}\odot\textbf{B}=
[\textbf{a}_1\otimes \textbf{b}_1 \quad \textbf{a}_2\otimes \textbf{b}_2 \quad \cdots \quad \textbf{a}_K\otimes \textbf{b}_K]
\end{equation} 

\subsection{Tensor Decomposition}
In our approach, we use the CP decomposition introduced by Harshman in \cite{c5} and Carroll and Chang in \cite{c6}. This decomposition has the advantage of being one of the simplest tensor decomposition. It represents a tensor $\mathscr{X}\in\mathbb{R}^{I_1\times I_2\times ...\times I_N}$ as the sum of component of vector outer products. 
\begin{equation} \label{eq::cp}
\mathscr{X}=\sum_{r=1}^R \textbf{a}_r^{(1)} \circ \textbf{a}_r^{(2)} \circ \cdots \circ \textbf{a}_r^{(N)}
\end{equation}

\begin{figure}[b]
\begin{center}
\begin{tikzpicture}[scale=0.6, transform shape]
\draw (0,0) rectangle (2.5,3) node (v1) {};
\draw (2.5,3) -- (4.5,5);
\draw (0,3) -- (2,5);
\draw (2.5,0) -- (4.5,2);
\draw (2.0,5) -- (4.5,5);
\draw (4.5,2) -- (4.5,5);
\draw [dashdotted] (2.0,2) -- (4.5,2);
\draw [dashdotted] (2.0,2) -- (2.0,5);
\draw [dashdotted] (0,0) -- (2.0,2.0);

\draw (1.25,0) node[below]{\Large{$I_1$}};
\draw (0,1.5) node[left]{\Large{$I_2$}};
\draw (0.9,4.0) node[left]{\Large{$I_3$}};
\draw (5.5,3.5) node[left]{\Large{$=$}};

\draw [->] (6,3.5) -- (6.0,1);
\draw (6,2.75) node[left]{\Large{$a_1$}};
\draw [->] (6,3.5) -- (7.5,3.5);
\draw (6.75,3.5) node[below]{\Large{$b_1$}};
\draw [->] (6,3.5) -- (7.5,5.0);
\draw (6.5,4.0) node[left]{\Large{$c_1$}};

\draw (8.5,3.5) node[left]{\Large{$+$}};
\draw (9.0,3.5) node[left]{\Large{$...$}};
\draw (9.5,3.5) node[left]{\Large{$+$}};

\draw [->] (10,3.5) -- (10.0,1);
\draw (10,2.75) node[left]{\Large{$a_R$}};
\draw [->] (10,3.5) -- (11.5,3.5);
\draw (10.75,3.5) node[below]{\Large{$b_R$}};
\draw [->] (10,3.5) -- (11.5,5.0);
\draw (10.5,4.0) node[left]{\Large{$c_R$}};
\end{tikzpicture}
\caption{CANDECOMP/PARAFAC decomposition into R components of a three way tensor}
\end{center}
\end{figure}
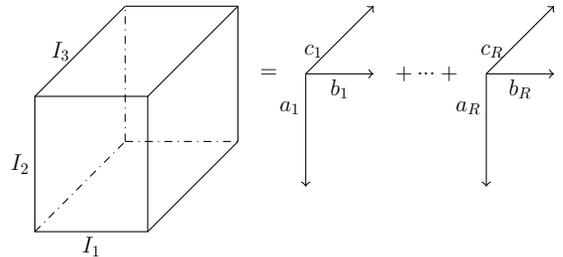
To achieve the computation of the CP decomposition, the following minimization equation has to be solved.
\begin{equation} \label{eq::minimztn}
\min_{\mathscr{\hat{X}}} ||\mathscr{X}-\mathscr{\hat{X}}||
\end{equation}
with $\mathscr{\hat{X}}$ the approximate tensor described by the CP decomposition and $\mathscr{X}$ the original tensor.

To solve equation \ref{eq::minimztn}, the Alternating Least  Squares (ALS) method is used as presented by Harshman in \cite{c5} and Carroll and Chang in \cite{c6}. In the experiments, we use the non-negative CP decomposition introduced by Lee and Seung in \cite{c13} for easier post-treatment. The matrices $\textbf{A}\in\mathbb{R}^{I\times R}$, $\textbf{B}\in\mathbb{R}^{J\times R}$ and $\textbf{C}\in\mathbb{R}^{K\times R}$ are now updated according to the multiplicative update rule for a tensor of size $\mathscr{X}\in\mathbb{R}^{I\times J \times K}$.
\begin{equation} \label{eq::als_upd}
\begin{cases}
a_{ir}\leftarrow& a_{ir} \dfrac{[\textbf{X}_{(1)}(\textbf{C}\odot\textbf{B})]_{ir}}{[\textbf{A}(\textbf{C}\odot\textbf{B})^T(\textbf{C}\odot\textbf{B})]_{ir}}\\
b_{jr}\leftarrow& b_{jr} \dfrac{[\textbf{X}_{(2)}(\textbf{C}\odot\textbf{A})]_{jr}}{[\textbf{B}(\textbf{C}\odot\textbf{A})^T(\textbf{C}\odot\textbf{A})]_{jr}}\\
c_{kr}\leftarrow& c_{kr} \dfrac{[\textbf{X}_{(3)}(\textbf{B}\odot\textbf{A})]_{kr}}{[\textbf{C}(\textbf{B}\odot\textbf{A})^T(\textbf{B}\odot\textbf{A})]_{kr}}
\end{cases}
\end{equation}
The multiplicative update rule helps to better calibration of the stochastic processes that uses the components of the tensor decomposition as a starting point. 

\section{STOCHASTIC SERIES PREDICTION}
In this section, we present first separately log-normal and mean-reverting stochastic models and then, we propose our approach consisting in a log-normal-mean-reverting stochastic model used for series prediction on smart contracts activities.

\subsection{Log-Normal Stochastic Diffusion Process}
The log-normal stochastic diffusion model, also known as geometric Brownian motion, is a continuous-time stochastic process. It is the solution of one of the most popular model in finance, the Black-Scholes model, introduced by Black and Scholes in \cite{c14}.

The model describes the evolution of a stock which is supposed to have a log-normal distribution of its returns. The stochastic process $S$ with a constant drift $\mu\in\mathbb{R}$, a constant volatility $\sigma\in\mathbb{R}$ and a Wiener process $W$ follows a geometric Brownian motion if the following equation is satisfied.
\begin{equation} \label{eq::GBM}
dS_t=S_t(\mu dt+\sigma dW_t)
\end{equation}
The Wiener process, or Brownian motion, denoted by $W$ was introduced by R. Brown in \cite{c15} and represents the random motion of a small particle immersed in a fluid with the same density as the particle.

\subsection{Mean Reverting Stochastic Diffusion Process}
A mean-reverting process, also known as Ornstein-Uhlenbeck process, is a stochastic process that describes the velocity of a Brownian particle under friction. The process tends to evolve towards a specific long-term mean and it has been introduced by Ornstein and Uhlenbeck in \cite{c16}. This process was also generalized by Vasicek in \cite{c17} for wider application, especially in finance.

The stochastic process $r$ with a mean reversion speed $\lambda\in\mathbb{R}$, a long term mean $\kappa\in\mathbb{R}$, a volatility $\sigma\in\mathbb{R}$ and a Wiener process $W$ satisfies the following stochastic differential equation.
\begin{equation} \label{eq::MeanRvrtg}
dr_t=\lambda(\kappa-r_t)dt+\sigma dW_t
\end{equation} 

\subsection{Log-Normal-Mean-Reverting Model}
Our approach for the series modeling consists in the use of both the Ornstein-Uhlenbeck process and the geometric Brownian motion. The rationale is if a time series follows a log-normal distribution, it could be modeled according to the geometric Brownian motion model. On one side, volatility could be calibrated on the past evolution of the time series. On the other side, the drift should represent long term behavior if there is no volatility in the data set. In our log-normal-mean-reverting model, the drift is modeled with the Ornstein-Uhlenbeck process. Let define $S$ as the stochastic series process, $\mu$ as the stochastic drift process, $\sigma^{(S)}, \sigma^{(\mu)}\in\mathbb{R}$ the series volatility and the drift volatility, $\lambda, \kappa\in\mathbb{R}$ the mean-reversion speed and the long term mean, $W$ as a Brownian motion and $\rho$ as the correlation. Our model is defined by the system of equations below.

\begin{equation} \label{eq::FinMdl}
\begin{cases}
dS_t=S_t(\mu_t dt+\sigma^{(S)} dW_t^{(1)}) &  \\ 
d\mu_t=\lambda(\kappa-\mu_t)dt+\sigma^{(\mu)} dW_t^{(2)} &  \\ 
\rho dt=d \langle W^{(1)},W^{(2)} \rangle_t & 
\end{cases}
\end{equation}

The correlation denoted by $\rho$ characterizes the correlation between the two Brownian motions of the Geometric Brownian Motion and the Ornstein-Uhlenbeck process, denoted respectively by $W^{(1)}$ and $W^{(2)}$. 

\section{EXPERIMENTS}
In this section, we describe the data used for the tensor decomposition and the simulation of smart contracts activities using our log-normal-mean-reverting model with the goal of speculative investment.\\
All the experiments are performed on a PC with Intel Core i7 CPU and 8 GB of RAM. The algorithm for non-negative CP decomposition and stochastic processes has been implemented in Python language.

\subsection{Data from Smart Contracts and Tensor Completion}
Smart contracts data have been collected from the Ethereum platform starting 7 August 2015 and ending 2 March 2016. Different fields have been gathered such as hash key, sender accounts, receiver accounts, amount of Ether exchanged per transaction between two accounts and block heights. For the period considered within the data set, two millions transactions have been recorded. The average amount per transaction is approximately 76 Ethers. The average number of transactions per sender account is 47 transactions and per receiver account is 26 transactions.

A three-way tensor is defined according to the smart contracts data. The first dimension of the tensor, $I$, represents the sender accounts, the second dimension of the tensor, $J$, the receiver accounts and the third dimension, $K$, the time slot. The interaction at a given time slot between a sender account and a receiver account is represented by the amount of Ether exchanged. 

\subsection{Selection of the Smart Contracts Data For Tensor Decomposition}
Among the data collected from the Ethereum platform, around 60\% of the sender contracts send only one payment. That is around 25,000 contracts. Around 70\% of the contracts, 50,000 contracts, receive only one payment for the time period considered. To concentrate more on regular smart contract activities, we decide to consider the 1\% most active contracts during the  time.
The resulting tensor has a size of 459$\times$813$\times$52. 

\begin{figure}[b]
\begin{center}
\begin{tikzpicture}[scale=0.40, transform shape]
\draw (0,0) -- (0,7);
\draw (0,7) -- (2.7,7);
\draw (0,0) -- (7,0);
\draw (7,0) -- (7,3.7);
\draw (7,0) -- (10.4,5.4);
\draw (10.4,5.4) -- (10.4,12.4);
\draw (4.4,12.4) -- (10.4,12.4);
\draw (0,7) -- (4.4,12.4);
\draw [dashdotted] (4.4,8.7) -- (4.4,12.4);
\draw [dashdotted] (7.7,5.4) -- (10.4,5.4);

\draw (0.5,0.5) rectangle (5.5,5.5);
\draw (1.5,0.5) -- (1.5,5.5);
\draw (2.5,0.5) -- (2.5,5.5);
\draw (3.5,0.5) -- (3.5,5.5);
\draw (4.5,0.5) -- (4.5,5.5);
\draw (0.5,1.5) -- (5.5,1.5);
\draw (0.5,2.5) -- (5.5,2.5);
\draw (0.5,3.5) -- (5.5,3.5);
\draw (0.5,4.5) -- (5.5,4.5);
\draw (0.5,4) node[right]{\Large{$Se_1$}};
\draw (0.5,3) node[right]{\Large{$Se_2$}};
\draw (0.5,2) node[right]{\Large{$Se_3$}};
\draw (0.8,1) node[right]{\Large{$\vdots$}};
\draw (1.8,1) node[right]{\Large{$\vdots$}};
\draw (2.8,1) node[right]{\Large{$\vdots$}};
\draw (3.8,1) node[right]{\Large{$\vdots$}};
\draw (4.6,1) node[right]{\Large{$\ddots$}};
\draw (1.5,5) node[right]{\Large{$Re_1$}};
\draw (2.5,5) node[right]{\Large{$Re_2$}};
\draw (3.5,5) node[right]{\Large{$Re_3$}};
\draw (4.57,5) node[right]{\Large{$\cdots$}};
\draw (4.57,4) node[right]{\Large{$\cdots$}};
\draw (4.57,3) node[right]{\Large{$\cdots$}};
\draw (4.57,2) node[right]{\Large{$\cdots$}};
\draw (1.75,4) node[right]{\Large{1}};
\draw (2.65,4) node[right]{\Large{15}};
\draw (3.75,4) node[right]{\Large{0}};
\draw (1.75,3) node[right]{\Large{0}};
\draw (1.75,2) node[right]{\Large{5}};
\draw (2.75,2) node[right]{\Large{7}};
\draw (2.65,3) node[right]{\Large{30}};
\draw (3.75,2) node[right]{\Large{2}};
\draw (3.75,3) node[right]{\Large{0}};

\draw (2.7,8.7) -- (7.7,8.7);
\draw (2.7,5.5) -- (2.7,8.7);
\draw (3.7,5.5) -- (3.7,8.7);
\draw (4.7,5.5) -- (4.7,8.7);
\draw (5.7,3.7) -- (5.7,8.7);
\draw (6.7,3.7) -- (6.7,8.7);
\draw (7.7,3.7) -- (7.7,8.7);
\draw (2.7,7.7) -- (7.7,7.7);
\draw (2.7,6.7) -- (7.7,6.7);
\draw (2.7,5.7) -- (7.7,5.7);
\draw (5.5,4.7) -- (7.7,4.7);
\draw (5.5,3.7) -- (7.7,3.7);
\draw (2.7,6.2) node[right]{\Large{$Se_2$}};
\draw (2.7,7.2) node[right]{\Large{$Se_1$}};
\draw (3.7,8.2) node[right]{\Large{$Re_1$}};
\draw (4.7,8.2) node[right]{\Large{$Re_2$}};
\draw (5.7,8.2) node[right]{\Large{$Re_3$}};
\draw (6.7,8.2) node[right]{\Large{$\cdots$}};
\draw (6.7,7.2) node[right]{\Large{$\cdots$}};
\draw (6.7,6.2) node[right]{\Large{$\cdots$}};
\draw (6.7,5.2) node[right]{\Large{$\cdots$}};
\draw (6.8,4.2) node[right]{\Large{$\ddots$}};
\draw (3.95,7.2) node[right]{\Large{1}};
\draw (4.85,7.2) node[right]{\Large{12}};
\draw (5.95,7.2) node[right]{\Large{0}};
\draw (3.95,6.2) node[right]{\Large{3}};
\draw (4.85,6.2) node[right]{\Large{28}};
\draw (5.95,6.2) node[right]{\Large{0}};
\draw (5.95,5.2) node[right]{\Large{0}};
\draw (5.95,4.2) node[right]{\Large{$\vdots$}};

\draw (4.9,9.9) -- (9.9,9.9);
\draw (4.9,10.9) -- (9.9,10.9);
\draw (4.9,11.9) -- (9.9,11.9);
\draw (4.9,8.9) -- (9.9,8.9);
\draw (7.7,7.9) -- (9.9,7.9);
\draw (7.7,6.9) -- (9.9,6.9);
\draw (4.9,8.7) -- (4.9,11.9);
\draw (5.9,8.7) -- (5.9,11.9);
\draw (6.9,8.7) -- (6.9,11.9);
\draw (7.9,6.9) -- (7.9,11.9);
\draw (8.9,6.9) -- (8.9,11.9);
\draw (9.9,6.9) -- (9.9,11.9);
\draw (4.9,10.4) node[right]{\Large{$Se_1$}};
\draw (4.9,9.4) node[right]{\Large{$Se_2$}};
\draw (5.9,11.4) node[right]{\Large{$Re_1$}};
\draw (6.9,11.4) node[right]{\Large{$Re_2$}};
\draw (7.9,11.4) node[right]{\Large{$Re_3$}};
\draw (9,11.4) node[right]{\Large{$\cdots$}};
\draw (9,10.4) node[right]{\Large{$\cdots$}};
\draw (9,9.4) node[right]{\Large{$\cdots$}};
\draw (9,8.4) node[right]{\Large{$\cdots$}};
\draw (9,7.4) node[right]{\Large{$\ddots$}};
\draw (8.15,7.4) node[right]{\Large{$\vdots$}};
\draw (5.95,10.4) node[right]{\Large{2}};
\draw (7.05,10.4) node[right]{\Large{13}};
\draw (8.15,10.4) node[right]{\Large{1}};
\draw (6.15,9.4) node[right]{\Large{5}};
\draw (7.05,9.4) node[right]{\Large{26}};
\draw (8.15,8.4) node[right]{\Large{2}};
\draw (8.15,9.4) node[right]{\Large{2}};

\draw (10.0,11.9) -- (10.4,12.4);

\draw (3.5,0) node[below]{\Large{Receiver}};
\draw (0,3.05) node[above, rotate=90]{\Large{Sender}};
\draw (9.25,2.5) node[above, rotate=60]{\Large{Time}};

\fill [gray, opacity=0.2] (0.5,0.5) rectangle (5.5,5.5);
\fill [gray, opacity=0.2] (2.7,3.7) rectangle (7.7,8.7);
\fill [gray, opacity=0.2] (4.9,6.9) rectangle (9.9,11.9);
\end{tikzpicture}
\caption{Three-way tensor containing Ether amount exchanged between different sender and receiver accounts}
\end{center}
\end{figure}
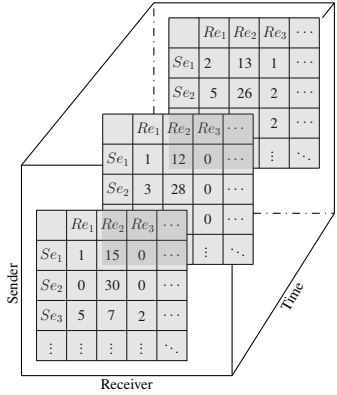

\begin{figure}[!b]
	\begin{center}
	\includegraphics[scale=0.40]{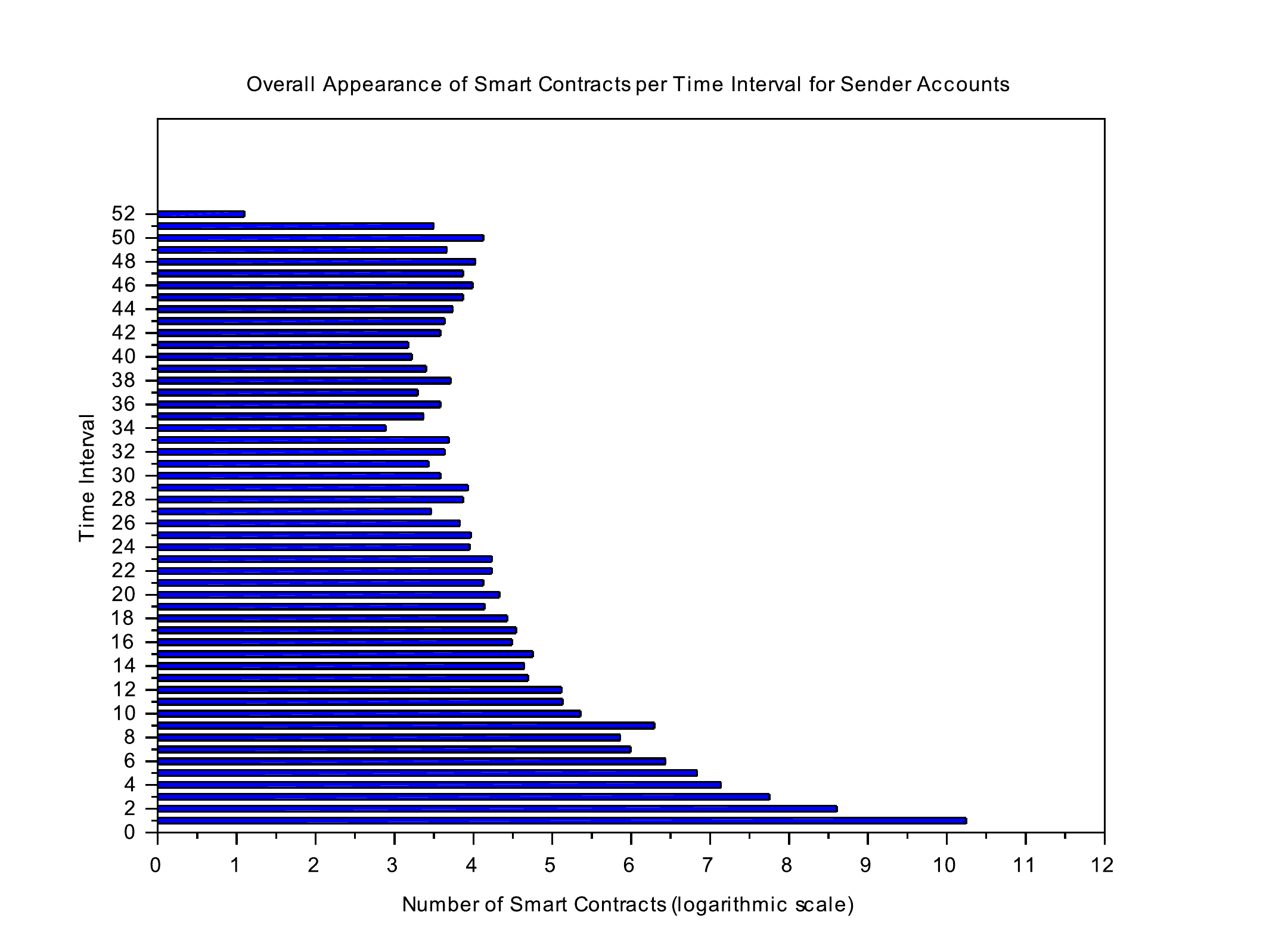}
	\caption{More than 25,000 smart contracts only received one transaction in the  time frame.}
	\end{center}
\end{figure}

\begin{figure}[!b]
	\begin{center}
	\includegraphics[scale=0.40]{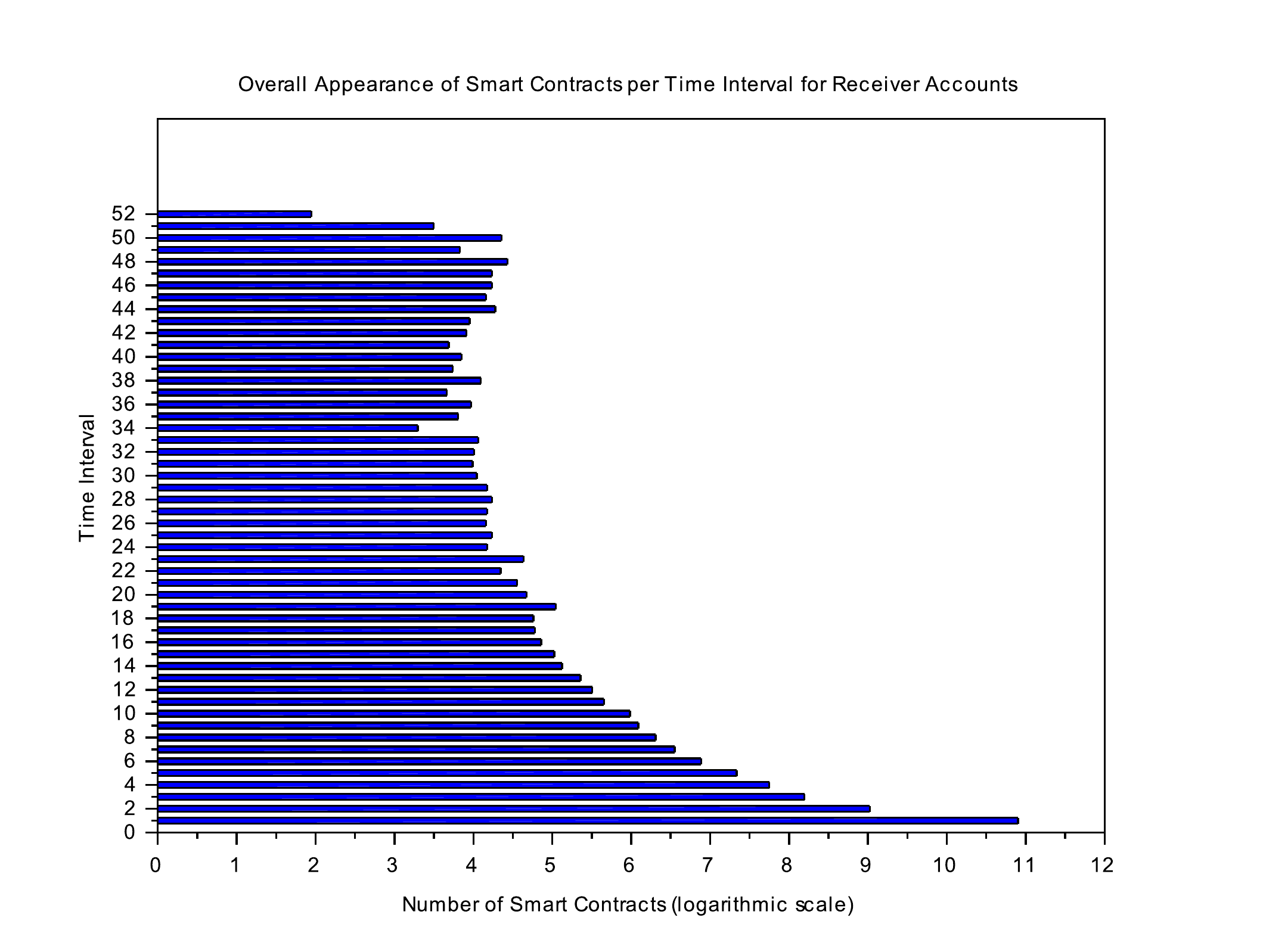}
	\caption{More than 50,000 smart contracts only received one transaction during the  time frame.}
	\end{center}
\end{figure} 

\subsection{Application of the Non-Negative CP Decomposition}
Non-negative CP decomposition is applied to the smart contracts tensor. The choice of the use of a non-negative algorithm is mainly for easier calibration of the stochastic processes on the tensor decomposition components. 

We define a stopping criterion $\epsilon$ for ALS algorithm using the evolution in the norm of the approximate tensor.
\begin{equation} \label{eq::cvrgc}
||\mathscr{\hat{X}}||^{k^{th}step}-||\mathscr{\hat{X}}||^{(k-1)^{th}step}\leq\epsilon \quad , \quad \epsilon=0.001
\end{equation}
We estimate a number of rank equals to five for the tensor decomposition as the data observed within the data set can be decomposed as small exchanges, moderate exchanges, active exchanges and very active exchanges. According to the rank, the tensor decomposition highlights the interactions between senders and receivers in function of time. 
In figure 5, one sender account has been selected to visualize the Ether amount exchanged with different receiver accounts based on CP decomposition. 

Furthermore, numerical experience shows that the description $\frac{\text{factor time t}}{\text{factor time t-1}}$ for a specified rank follows a log-normal distribution. To assess the accuracy of the fit to log-normal distribution, we perform the Shapiro normal test, as a distribution is said to be log-normal if the natural logarithm of the distribution is normally distributed. For our data set, we define a p-value of 10\% for the null hypothesis that is the data set follows a log-normal distribution. The results are presented in table 1. 

\begin{table}[!b]
\begin{center}
\caption{Results of Shapiro log-normal test}
\label{table::Shapiro}
\scalebox{1.1}{
\begin{tabular}{|c|c|c|c|c|c|}
\hline
Rank & 1 & 2 & 3 & 4 & 5\\
\hline
pValue & 0.1298 & 0.0003 & 0.0029 & 0.0905 & 0.0003 \\
\hline
\end{tabular}}
\end{center}
\end{table}

It can be observed that the p-value of the first rank is just outside the threshold of 10\%. However, we decide that the stochastic processes described in equation \ref{eq::FinMdl} would still describe properly the series for tensor rank 1 as the p-value is equal to 12.98\%.  

\subsection{Use Of The Log-Normal-Mean-Reverting Process}
Our time series consists in fifty-two time events. The calibration of the process $S_t$ is performed historically using the first twenty-six time events for the simulation of the next twenty-six events, the first forty-two time events for the simulation of the next ten events and the first forty-seven time events for the simulation of the next five events. The prediction is then analyzed with the original data of the same time period to validate the approach.

\begin{figure}[!b]
	\begin{center}
	\includegraphics[scale=0.40]{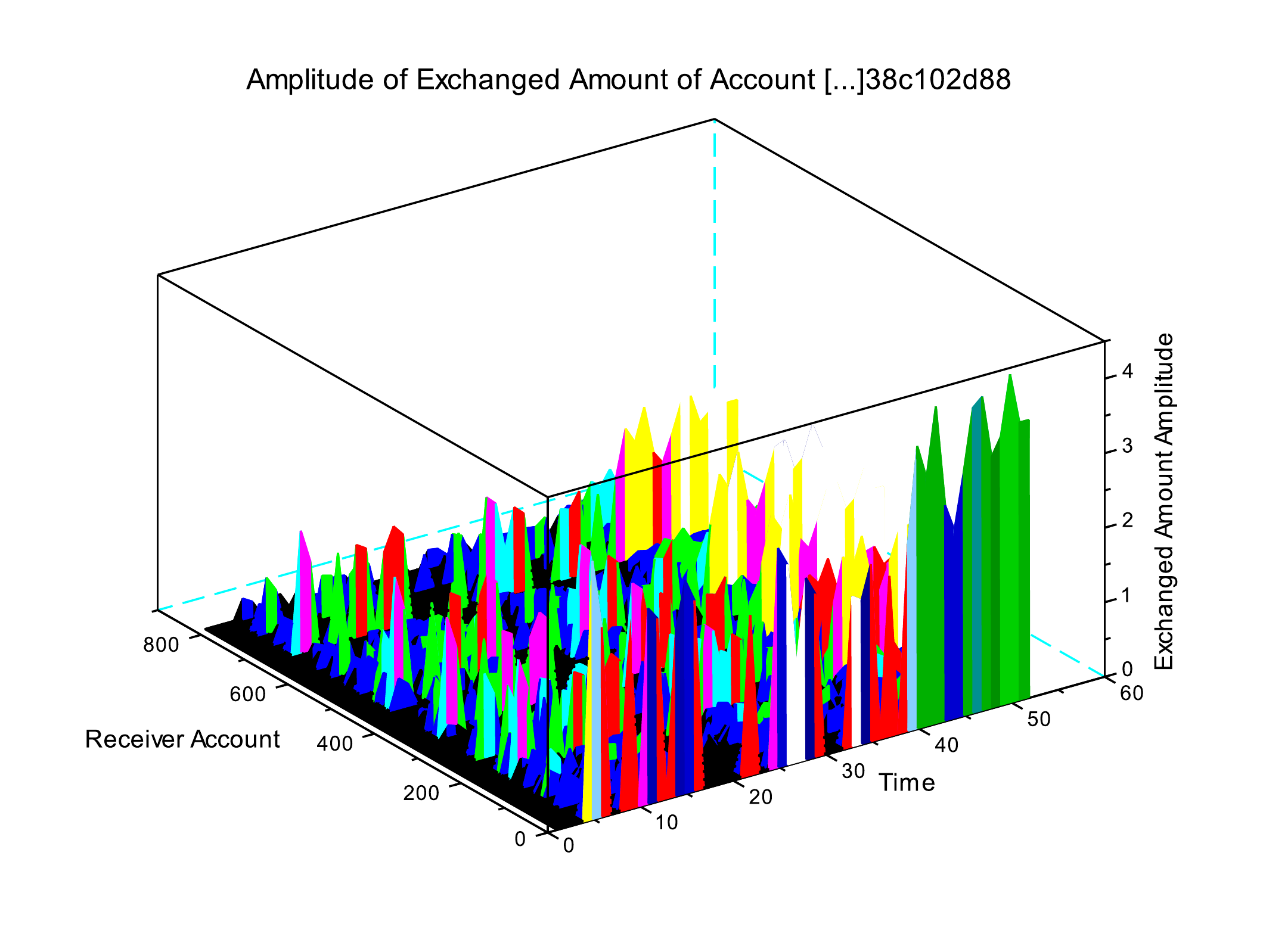}
	\caption{Amplitude of exchanged Ether amount from a given sender to receivers during the time interval}
	\end{center}
\end{figure}

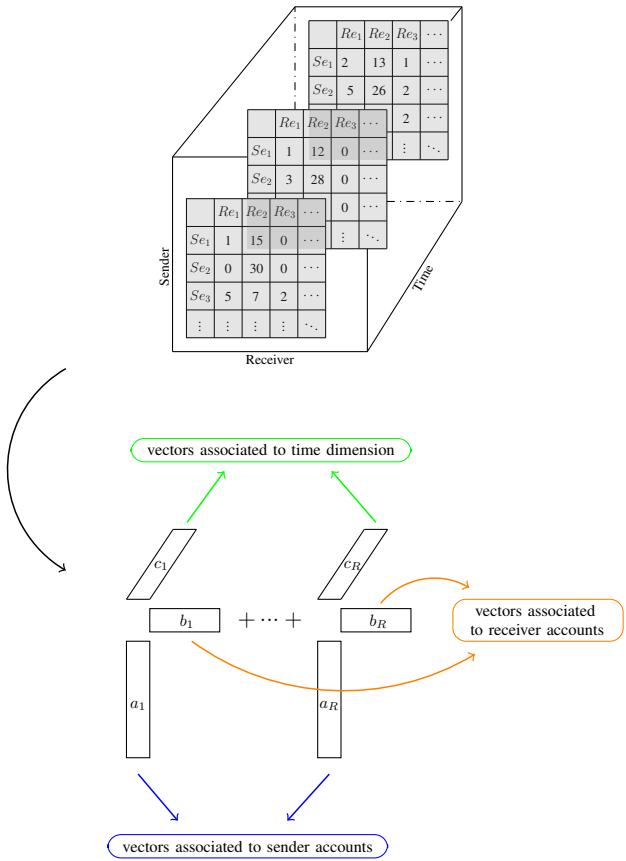
\begin{figure}[!b]
\begin{center}
\begin{tikzpicture}[scale=0.37, transform shape]
\draw (0,10.6) -- (0,17.6);
\draw (0,17.6) -- (2.7,17.6);
\draw (0,10.6) -- (7,10.6);
\draw (7,10.6) -- (7,14.3);
\draw (7,10.6) -- (10.4,16);
\draw (10.4,16) -- (10.4,23);
\draw (4.4,23) -- (10.4,23);
\draw (0,17.6) -- (4.4,23);
\draw [dashdotted] (4.4,19.3) -- (4.4,23);
\draw [dashdotted] (7.7,16) -- (10.4,16);

\draw (0.5,11.1) rectangle (5.5,16.1);
\draw (1.5,11.1) -- (1.5,16.1);
\draw (2.5,11.1) -- (2.5,16.1);
\draw (3.5,11.1) -- (3.5,16.1);
\draw (4.5,11.1) -- (4.5,16.1);
\draw (0.5,12.1) -- (5.5,12.1);
\draw (0.5,13.1) -- (5.5,13.1);
\draw (0.5,14.1) -- (5.5,14.1);
\draw (0.5,15.1) -- (5.5,15.1);
\draw (0.5,14.6) node[right]{\Large{$Se_1$}};
\draw (0.5,13.6) node[right]{\Large{$Se_2$}};
\draw (0.5,12.6) node[right]{\Large{$Se_3$}};
\draw (0.8,11.6) node[right]{\Large{$\vdots$}};
\draw (1.8,11.6) node[right]{\Large{$\vdots$}};
\draw (2.8,11.6) node[right]{\Large{$\vdots$}};
\draw (3.8,11.6) node[right]{\Large{$\vdots$}};
\draw (4.6,11.6) node[right]{\Large{$\ddots$}};
\draw (1.5,15.6) node[right]{\Large{$Re_1$}};
\draw (2.5,15.6) node[right]{\Large{$Re_2$}};
\draw (3.5,15.6) node[right]{\Large{$Re_3$}};
\draw (4.57,15.6) node[right]{\Large{$\cdots$}};
\draw (4.57,14.6) node[right]{\Large{$\cdots$}};
\draw (4.57,13.6) node[right]{\Large{$\cdots$}};
\draw (4.57,12.6) node[right]{\Large{$\cdots$}};
\draw (1.75,14.6) node[right]{\Large{1}};
\draw (2.65,14.6) node[right]{\Large{15}};
\draw (3.75,14.6) node[right]{\Large{0}};
\draw (1.75,13.6) node[right]{\Large{0}};
\draw (1.75,12.6) node[right]{\Large{5}};
\draw (2.75,12.6) node[right]{\Large{7}};
\draw (2.65,13.6) node[right]{\Large{30}};
\draw (3.75,12.6) node[right]{\Large{2}};
\draw (3.75,13.6) node[right]{\Large{0}};

\draw (2.7,19.3) -- (7.7,19.3);
\draw (2.7,16.1) -- (2.7,19.3);
\draw (3.7,16.1) -- (3.7,19.3);
\draw (4.7,16.1) -- (4.7,19.3);
\draw (5.7,14.3) -- (5.7,19.3);
\draw (6.7,14.3) -- (6.7,19.3);
\draw (7.7,14.3) -- (7.7,19.3);
\draw (2.7,18.3) -- (7.7,18.3);
\draw (2.7,17.3) -- (7.7,17.3);
\draw (2.7,16.3) -- (7.7,16.3);
\draw (5.5,15.3) -- (7.7,15.3);
\draw (5.5,14.3) -- (7.7,14.3);
\draw (2.7,16.8) node[right]{\Large{$Se_2$}};
\draw (2.7,17.8) node[right]{\Large{$Se_1$}};
\draw (3.7,18.8) node[right]{\Large{$Re_1$}};
\draw (4.7,18.8) node[right]{\Large{$Re_2$}};
\draw (5.7,18.8) node[right]{\Large{$Re_3$}};
\draw (6.7,18.8) node[right]{\Large{$\cdots$}};
\draw (6.7,17.8) node[right]{\Large{$\cdots$}};
\draw (6.7,16.8) node[right]{\Large{$\cdots$}};
\draw (6.7,15.8) node[right]{\Large{$\cdots$}};
\draw (6.8,14.8) node[right]{\Large{$\ddots$}};
\draw (3.95,17.8) node[right]{\Large{1}};
\draw (4.85,17.8) node[right]{\Large{12}};
\draw (5.95,17.8) node[right]{\Large{0}};
\draw (3.95,16.8) node[right]{\Large{3}};
\draw (4.85,16.8) node[right]{\Large{28}};
\draw (5.95,16.8) node[right]{\Large{0}};
\draw (5.95,15.8) node[right]{\Large{0}};
\draw (5.95,14.8) node[right]{\Large{$\vdots$}};

\draw (4.9,20.5) -- (9.9,20.5);
\draw (4.9,21.5) -- (9.9,21.5);
\draw (4.9,22.5) -- (9.9,22.5);
\draw (4.9,19.5) -- (9.9,19.5);
\draw (7.7,18.5) -- (9.9,18.5);
\draw (7.7,17.5) -- (9.9,17.5);
\draw (4.9,19.3) -- (4.9,22.5);
\draw (5.9,19.3) -- (5.9,22.5);
\draw (6.9,19.3) -- (6.9,22.5);
\draw (7.9,17.5) -- (7.9,22.5);
\draw (8.9,17.5) -- (8.9,22.5);
\draw (9.9,17.5) -- (9.9,22.5);
\draw (4.9,21) node[right]{\Large{$Se_1$}};
\draw (4.9,20) node[right]{\Large{$Se_2$}};
\draw (5.9,22) node[right]{\Large{$Re_1$}};
\draw (6.9,22) node[right]{\Large{$Re_2$}};
\draw (7.9,22) node[right]{\Large{$Re_3$}};
\draw (9,22) node[right]{\Large{$\cdots$}};
\draw (9,21) node[right]{\Large{$\cdots$}};
\draw (9,20) node[right]{\Large{$\cdots$}};
\draw (9,19) node[right]{\Large{$\cdots$}};
\draw (9,18) node[right]{\Large{$\ddots$}};
\draw (8.15,18) node[right]{\Large{$\vdots$}};
\draw (5.95,21) node[right]{\Large{2}};
\draw (7.05,21) node[right]{\Large{13}};
\draw (8.15,21) node[right]{\Large{1}};
\draw (6.15,20) node[right]{\Large{5}};
\draw (7.05,20) node[right]{\Large{26}};
\draw (8.15,19) node[right]{\Large{2}};
\draw (8.15,20) node[right]{\Large{2}};

\draw (10,22.5) -- (10.4,23);

\draw (3.5,10.6) node[below]{\Large{Receiver}};
\draw (0,13.65) node[above, rotate=90]{\Large{Sender}};
\draw (9.25,13.1) node[above, rotate=60]{\Large{Time}};

\fill [gray, opacity=0.2] (0.5,11.1) rectangle (5.5,16.1);
\fill [gray, opacity=0.2] (2.7,14.3) rectangle (7.7,19.3);
\fill [gray, opacity=0.2] (4.9,17.5) rectangle (9.9,22.5);
\end{tikzpicture}
\begin{tikzpicture}[scale=0.62, transform shape]
\draw (0.8,1.65) rectangle (1.3,4.15) node (v2) {};
\draw (1.05,3) node[below]{$a_1$};

\draw (1.3,4.35) rectangle (2.8,4.85) node (v3) {};
\draw (2.1,4.85) node[below]{$b_1$};

\draw (0.8,5.05) -- (1.3,5.05);
\draw (0.8,5.05) -- (1.8,6.55);
\draw (1.8,6.55) -- (2.3,6.55);
\draw (2.3,6.55) -- (1.3,5.05);
\draw (1.55,6) node[below]{$c_1$};

\draw (3.7,4.6) node[left]{\Large{$+$}};
\draw (4.2,4.6) node[left]{\Large{$...$}};
\draw (4.7,4.6) node[left]{\Large{$+$}};

\draw (4.9,1.65) rectangle (5.4,4.15) node (v4) {};
\draw (5.15,3) node[below]{$a_R$};

\draw (5.4,4.35) rectangle (6.9,4.85) node (v5) {};
\draw (6.2,4.85) node[below]{$b_R$};

\draw (5.4,5.05) -- (4.9,5.05);
\draw (5.4,5.05) -- (6.4,6.55);
\draw (5.9,6.55) -- (6.4,6.55);
\draw (5.9,6.55) -- (4.9,5.05);
\draw (5.65,6) node[below]{$c_R$};

\draw (3.4,0) node[below]{{vectors associated to sender accounts}};
\draw (8.15,4.8) node[right]{{vectors associated}};
\draw (8.05,4.4) node[right]{{to receiver accounts}};
\draw (3.9,8.5) node[below]{{vectors associated to time dimension}};

\draw (0.4,-0.5) [draw=blue, rounded corners=5pt] rectangle (6.4,0) node (v6) {}; 
\draw (0.9,8) [draw=green,rounded corners=5pt] rectangle (6.9,8.5) node (v7) {}; 
\draw (7.8,5.05) [draw=orange, rounded corners=5pt] rectangle (11.45,4.15) node (v8) {}; 

\draw [draw=blue, line width=0.2mm, ->] (1.05,1.3) -- (1.9,0.3);
\draw [draw=blue, line width=0.2mm, ->] (5.15,1.3) -- (4.25,0.3);

\draw [draw=green, line width=0.2mm, ->] (2.1,6.7) -- (2.9,7.8);
\draw [draw=green, line width=0.2mm, ->] (6.15,6.7) -- (5.2,7.8);

\draw [draw=orange, line width=0.2mm, ->] (2.2,4.15) arc (-126.1744:-59.0282:5.5);

\draw  [draw=orange, line width=0.2mm, ->]  (6.2509,4.9642) arc (140.0003:60:1.5);
\draw [draw=black, line width=0.2mm, ->] (-0.5,10) arc (120.0001:240:2.5);
\end{tikzpicture}
\caption{Description of the dimensions related to tensor decomposition. First dimension is linked to sender accounts, second dimension to receiver accounts and third dimension to time activity. Simulations are performed on the third dimension.}
\end{center}
\end{figure}

Using the system of equations described in \ref{eq::FinMdl}, six parameters have to be calibrated: the volatility of the series $\sigma^{(S)}$, the mean reverting speed and the long term mean, $\lambda$ and $\kappa$, the volatility $\sigma^{\mu}$ and the correlation $\rho$ between the two stochastic processes $S$ and $\mu$.

The volatility of the process $S_t$ is computed historically. The drift process $\mu_t$ illustrates the time value of money, also known as capitalization and actualization, that is one Ether today does not equal one Ether tomorrow. As a result, the parameters of the drift process $\mu_t$ are estimated on the Euro OverNight Index Average (EONIA) for the time period considered from 7 August 2015 to 2 March 2016. EONIA is the overnight rate exchanged in the interbank market. Due to the short time period of the Ether exchanged amount, it is more appropriate in this case to consider EONIA rates than other deposit rates with longer maturity. The last parameter, $\rho$ has to be calibrated before performing the series prediction. $\rho$ is the correlation between our time series extracted from the tensor decomposition and the EONIA historical rates. Exponential Weighted Moving Average (EWMA) correlation is used with a weight parameter of 0.9. 

\begin{table}[!b]
\begin{center}
\caption{Results of historical calibration of the stochastic processes}
\label{table::calibration}
\scalebox{1.1}{
\begin{tabular}{|c|c|c|c|c|c|}
\hline
Parameter & $\sigma^{(S)}$ & $\lambda$ & $ \kappa$ & $\sigma^{\mu}$ & $\rho$ \\
\hline
5 $\Delta$T & 0.5910 & 0.28180 & -0.0011 & 0.0000 &  -0.2621 \\
\hline
10 $\Delta$T &0.2010 & 0.2550 & -0.0011 & 0.0000 &  -0.2038 \\
\hline
26 $\Delta$T & 0.1672 & 0.1851 & -0.0011 & 0.0000 & -0.2288 \\
\hline
\end{tabular}}
\end{center}
\end{table}

The values of the parameters shown in table 2 are used for the time series predictions of five time steps, ten time steps and respectively twenty-six time steps. The Monte-Carlo method is chosen to solve the system of stochastic equations presented in \ref{eq::FinMdl} with one million simulation. 

\subsection{Selection Of The Smart Contracts for Speculative Investment}
The objective of the time series prediction using the stochastic processes is to evaluate the strength of the time vector for each tensor rank as described in figure 6. The selection of the smart contracts that exchange Ether is performed by assessing a probability for a time strength level. 

Each of the tensor rank is associated to a particular group of smart contracts as described in subsection 4.3. Each tensor rank highlights most relevant sender contracts related to receiver contracts according to a certain time frame. A larger value of amount exchanged between a sender and a receiver is characterized by a larger value in vector time in the tensor decomposition.

For the estimation of the future probabilities of the strength of the vector time for the different tensor ranks, a digital function is applied at the maturity of the log-normal-mean-reverting stochastic process. The digital function $C$ is defined by equation \ref{eq::dig}.
\begin{equation} \label{eq::dig}
C_T=\mathbbm{1}_{S_T\geq K}
\end{equation}

If the value of the stochastic process $S$ is below a level $K$ at maturity $T$, the value of $C_T$ is equal to 0. On the other hand, if the value of the stochastic process $S$ is higher or equal to a level $K$ at maturity $T$, the value of $C_T$ is equal to 1. This digital description allows to estimate the probability of the process $S$ to be higher or equal than a strike level $K$. The advantage of the use of the digital payoff is that the strike level can be defined according to the risk aversion of an investor. An investor having a risk averse profile would specify a high level of strike $K$ to maximize his probabilities of strong Ether exchanges even if it means that he might miss some opportunities. On the opposite, an investor having a risk taker profile would prefer to choose a lower strike value $K$ even if it means that sometimes the selected contracts won't receive Ether or could even have to send lot of Ether to other smart contracts.
Figure 7, 8 and 9 illustrate the relation between the risk that an investor is ready to take according to Ether exchange probabilities. Time series have been simulated for five time steps, ten time steps and twenty-six time steps. At each time step, the value of the digital is computed to retrieve the probability of Ether exchange. The probability can be either a receiving probability if a receiver account is selected or a sending probability if a sender account is selected. Finally, the probability is compared to the actual exchange of Ether in vector time. It is important to note that the payment probability gives a confidence value on the criteria that the series will be higher than a strike level. It can be seen as a reliable indicative measure for a speculative investment according to a risk profile or an investment strategy.
\begin{figure}[b!]
	\begin{center}
	\includegraphics[scale=0.40]{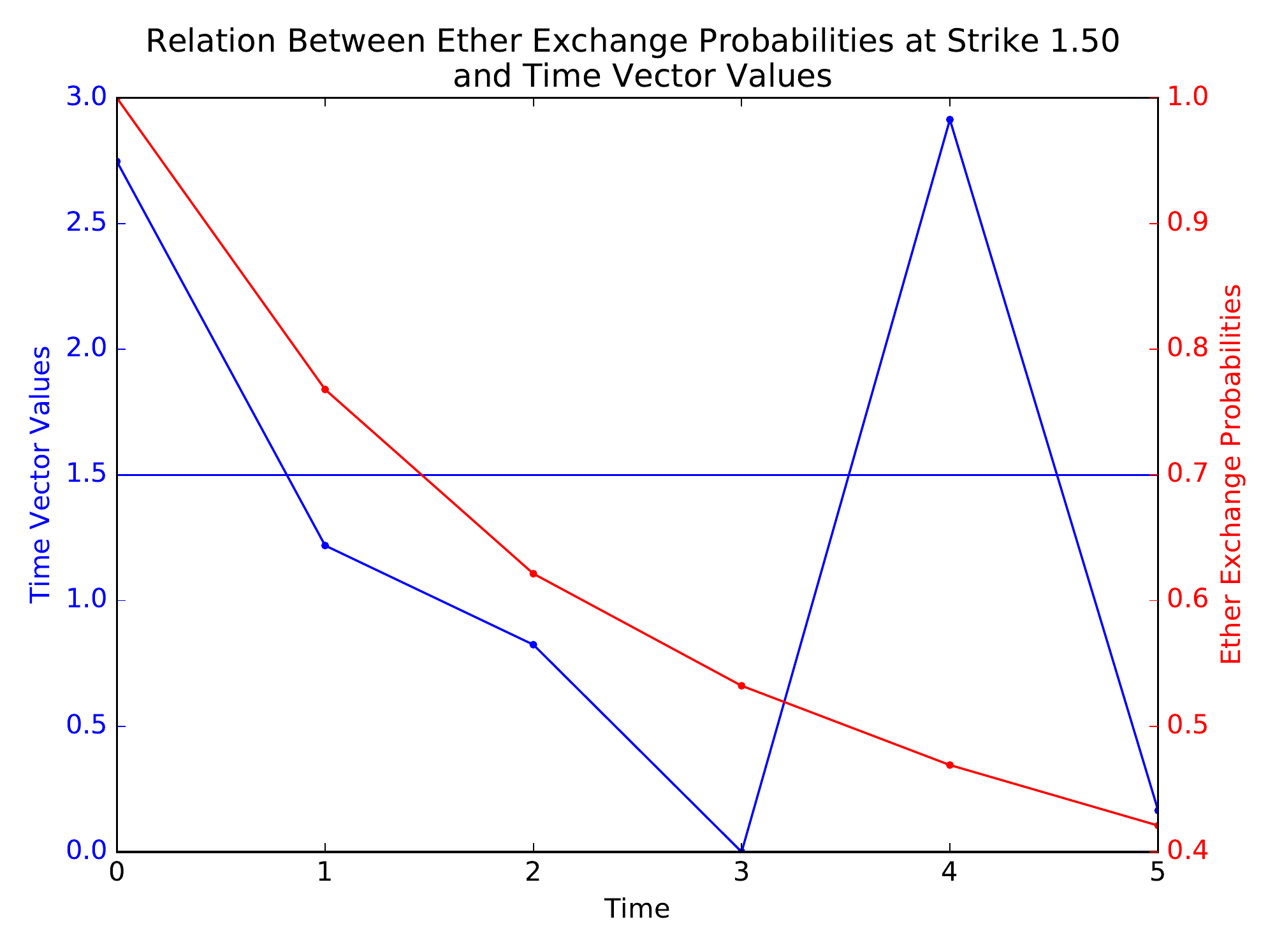}
	\caption{Relation between the time payment magnitude and the probability of receiving cash flows over time for 5 time steps.}
	\end{center}
\end{figure}
\begin{figure}[b!]
	\begin{center}
	\includegraphics[scale=0.40]{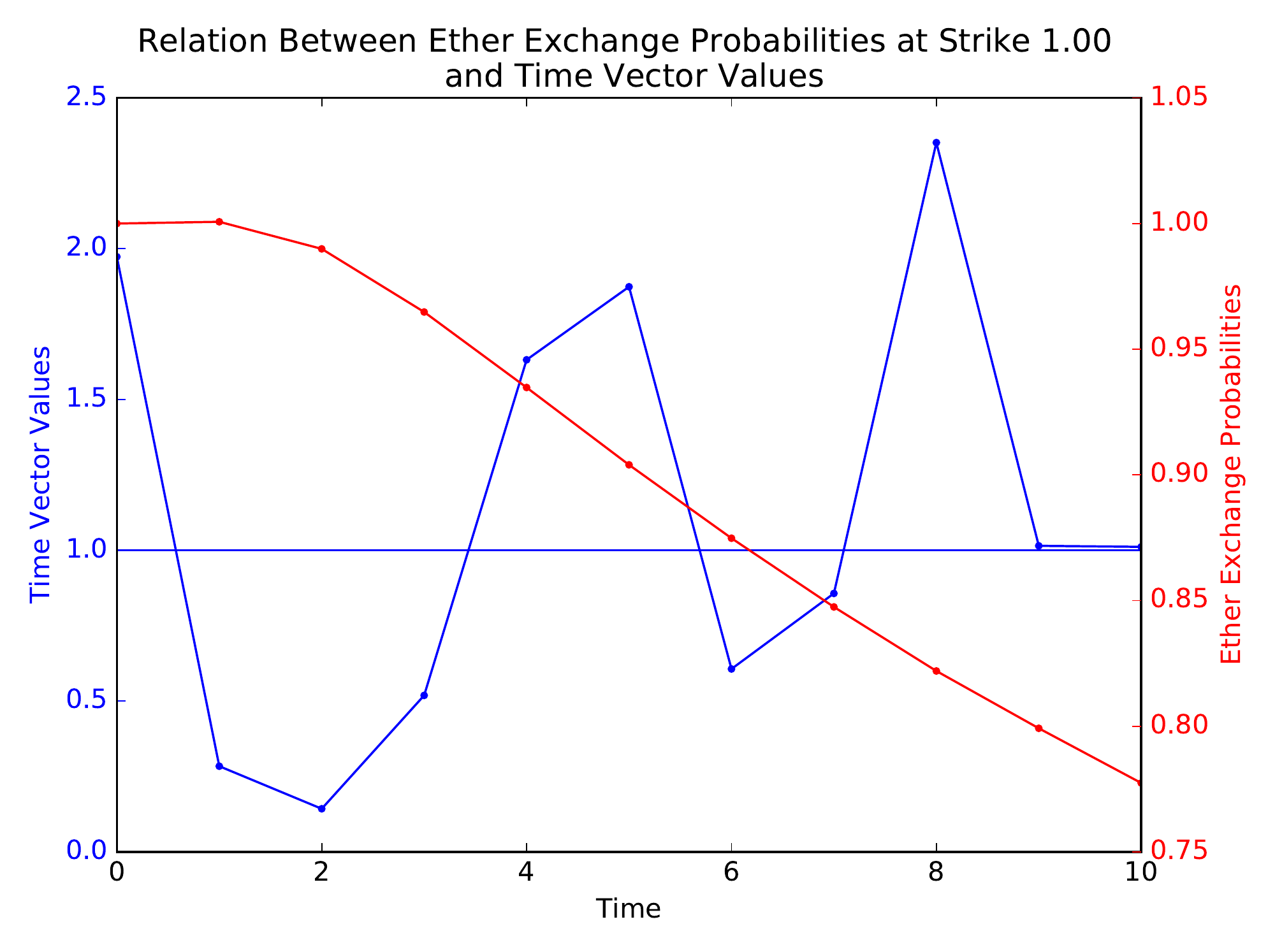}
	\caption{Relation between the time payment magnitude and the probability of receiving cash flows over time for 10 time steps.}
	\end{center}
\end{figure}
\begin{figure}[b]
	\begin{center}
	\includegraphics[scale=0.40]{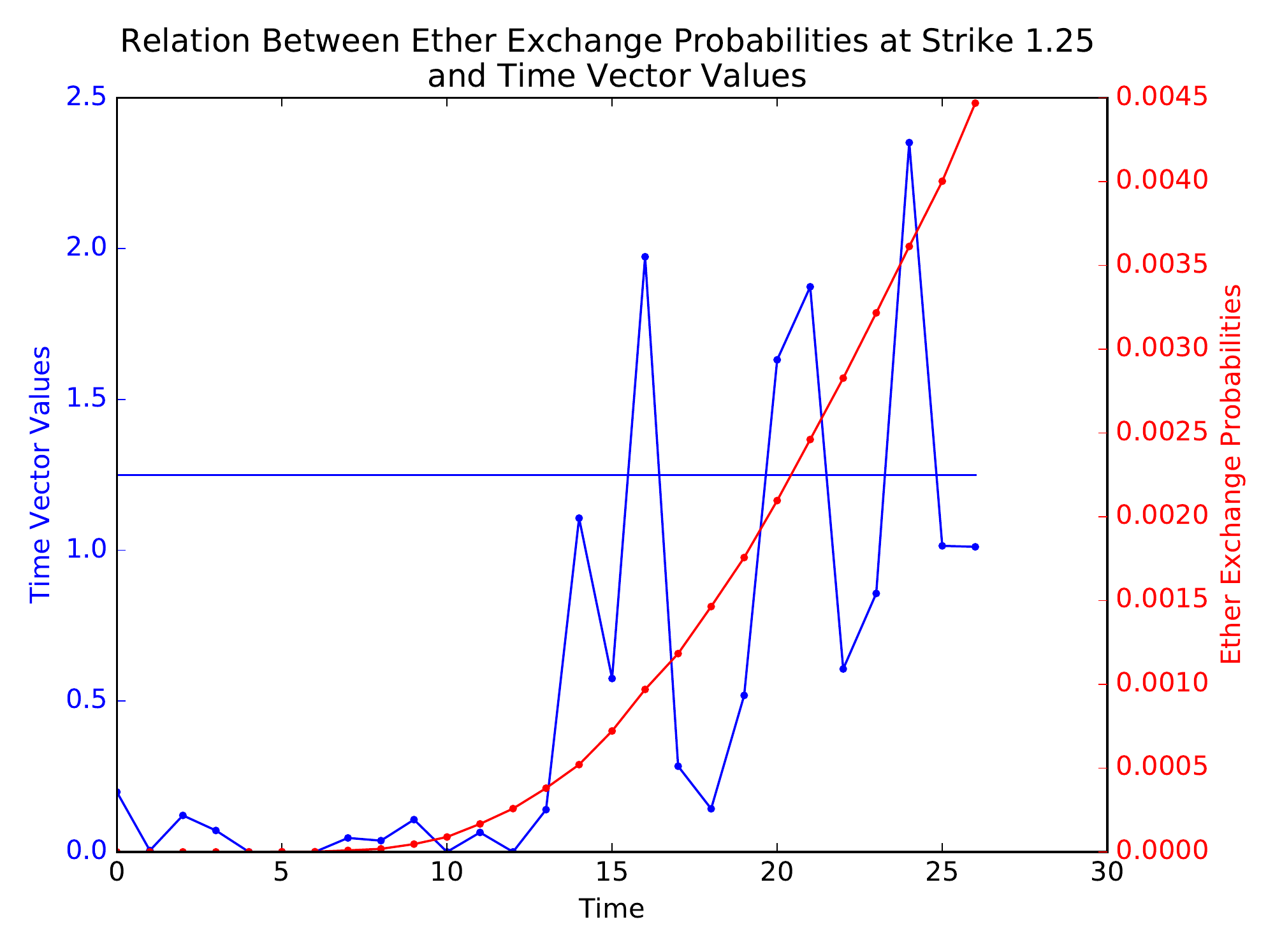}
	\caption{Relation between the time payment magnitude and the probability of receiving cash flows over time for 26 time steps.}
	\end{center}
\end{figure}

Tables 3, 4 and 5, the corresponding values of figures 7, 8 and 9, present the digital value in comparison to the actual value of the series for five time steps, ten time steps and twenty-six time steps of one tensor rank.
The digital value is strongly correlated to the time series values. Simulations lose accuracy when the time step is increasing as it introduces more uncertainty with longer simulated time. Digital value below 60\% reduces considerably the probabilities of exchanging Ether amount. In table 3, the probability value of 42\% means there are small probabilities of having a strong Ether exchange at fifth time step. Effectively, the time series is below the defined triggered level of 1.5. In table 4, at the tenth time step, there is a 70\% probability of exchanging Ether amount at a higher time level than 1.0. The actual value of the series confirms it with a value at the tenth time step of 1.0114. Similarly, in table 5, at the twenty-sixth time step, there is a 0.4\% probability of exchanging Ether amount at a higher strength level than 1.25 that is confirmed by the series value of 1.0114. To resume, the value of the digital can be considered as a strong indicator about the future exchanges in Ether. It provides a good source of information for speculative investment according to an investor-defined strength level of exchange in vector time.
\begin{table}[b!]
\begin{center}
\caption{Evolution of the digital value (Ether exchange probability) in relation with the series final value and the defined strength level $K=1.5$ for five time steps simulations}
\label{table::5tm_stp}
\scalebox{1.1}{
\begin{tabular}{|c|c|c|c|}
\hline
Time Step & Series Value & $\geq 1.5$ & Digital Value \\
\hline
0 & 2.7472 & - & - \\
5 & 0.1645 & 0 & 0.4218\\
\hline
\end{tabular}}
\end{center}
\end{table}

\begin{table}[b!]
\begin{center}
\caption{Evolution of the digital value (Ether exchange probability) in relation with the series evolution and the defined strength level $K=1.0$ for ten time steps simulations}
\label{table::10tm_stp}
\scalebox{1.1}{
\begin{tabular}{|c|c|c|c|}
\hline
Time Step & Series Value & $\geq 1.0$ & Digital Value \\
\hline
0 & 1.9732 & - & - \\
10 & 1.0114 & 1 & 0.7781 \\
\hline
\end{tabular}}
\end{center}
\end{table}

\begin{figure}[b!]
	\begin{center}
	\includegraphics[scale=0.40]{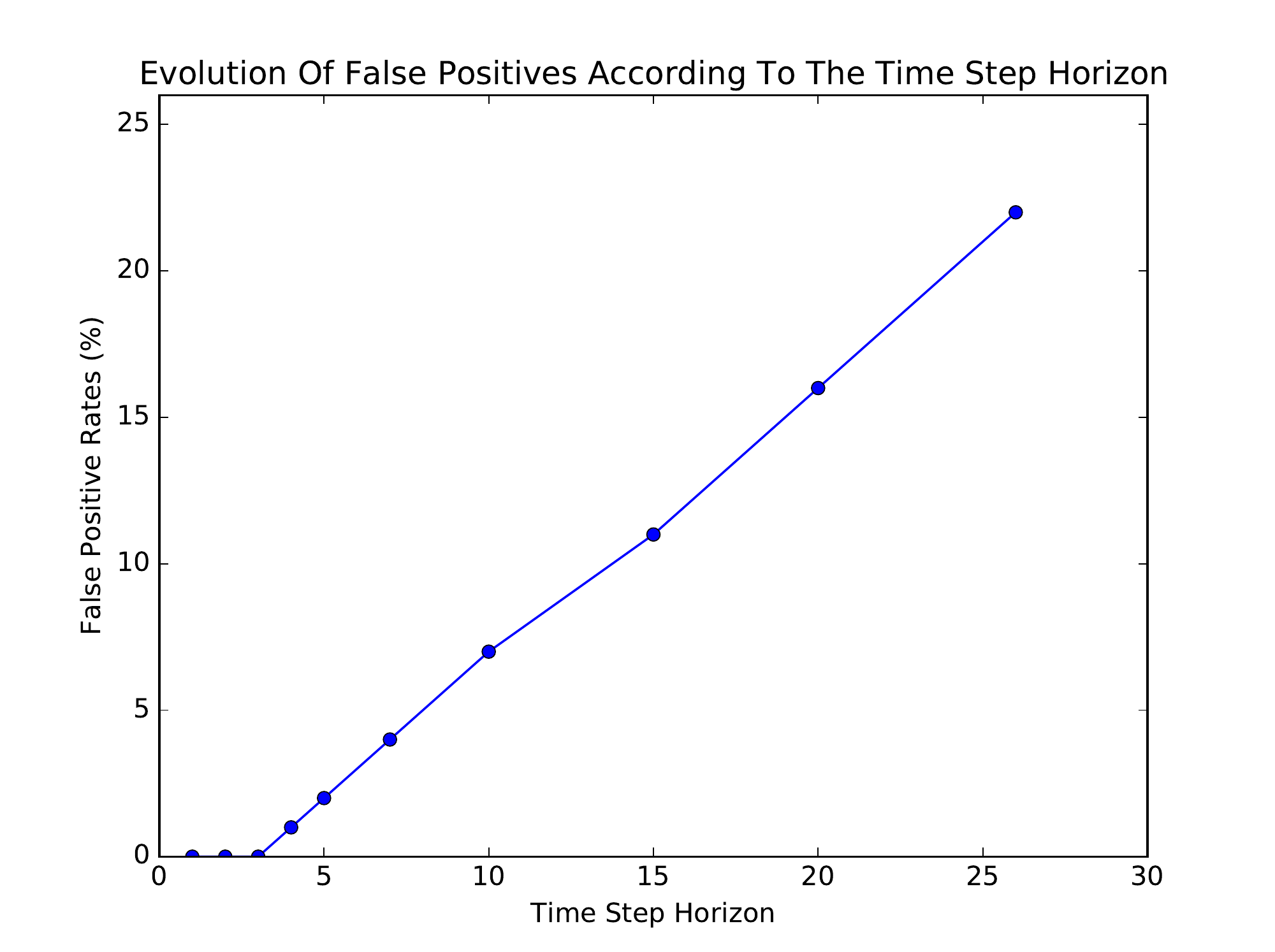}
	\caption{Relation between the time payment magnitude and the probability of receiving cash flows over time.}
    \label{fig::tfpositive}
	\end{center}
\end{figure}

Last but not least, the false positive and true positive rates have been calculated to determine the accuracy of the simulations. The results are shown in figure \ref{fig::tfpositive} and in table \ref{tab::tfpositive}. A false positive is defined when the probability of exchanging Ether is higher than 60\% for a strike level and no exchange of Ether happened or when the probability of exchanging Ether is below 60\% for a strike level and an exchange of Ether has been realized. Similarly, a true positive is defined either when the probability of exchange Ether is higher than 60\% according to a strike level and an exchange happened, or either when the probability of exchanging Ether was below the threshold of 60\% according to a strike level and no Ether exchange has been observed.

\begin{table}[b!]
\begin{center}
\caption{Evolution of the digital value (Ether exchange probability) in relation with the series evolution and the defined strength level $K=1.25$ for twenty-six time steps simulations}
\label{table::26tm_stp}
\scalebox{1.1}{
\begin{tabular}{|c|c|c|c|}
\hline
Time Step & Series Value & $\geq 1.25$ & Digital Value \\
\hline
0 & 0.1987 & - & - \\
26 & 1.0114 & 0 & 0.0045 \\
\hline
\end{tabular}}
\end{center}
\end{table}

\begin{table}[t]
\begin{center}
\caption{Evolution of the false positive and true positive rates depending on the horizon of the simulation for different strike levels} \label{tab::tfpositive}
\label{tbl::fptp}
\scalebox{1.1}{
\begin{tabular}{|c|c|c|}
\hline
Time Step & False Positive Rates (\%) & True Positive Rates (\%) \\
\hline
5 & 2 & 98 \\
10 & 7 & 93  \\
26 & 22 & 78  \\
\hline
\end{tabular}}
\end{center}
\end{table}


\section{CONCLUSIONS}
We address in this paper the problem of time series prediction applied to CP tensor decomposition using a stochastic process on smart contracts. We obtain accurate probabilities prediction of Ether exchange for sender and receiver accounts that could be fitted to the risk profile of an investor or to an investment strategy. As a result, our approach can be used for the analysis of smart contract activities but also for someone who is willing to consider smart contracts as a financial investment.

However, some  challenges will be  addressed in future work. One challenge is to use stochastic parameters for the volatility of the time series process or for the correlation involved in the stochastic equations system. It would help to increase accuracy of the simulations, in particular for longer time horizon, and to reflect deeper series variation over time. In addition, the well-known CP decomposition has been performed but other decomposition could be used to enrich the interaction analysis of the smart contracts activities such as the DEDICOM decomposition.




\end{document}